# Quantitative perfusion and water transport time model from multi b-value diffusion magnetic resonance imaging validated against neutron capture microspheres

| | | | |
|---|---|---|---|
| Mira Liu | Niloufar Saadat | Yong Jeong | Steven Roth |
| University of Chicago | University of Chicago | University of Chicago | University of Illinois |
| Medical Physics | Interventional Radiology | Medical Physics | Anesthesiology |
| Chicago, IL, USA | Chicago, IL, USA | Chicago, IL, USA | Chicago, IL, USA |
| liusarkarm@uchicago.edu | niloufar.saadat@gmail.com | yongjeong2013@u.northwestern.edu | rothgas@uic.edu |
| Marek Niekrasz | Mihai Giurcanu | Timothy Carroll | Gregory Christoforidis |
| University of Chicago | University of Chicago | University of Chicago | University of Chicago |
| Surgery | Statistics | Medical Physics | Interventional Radiology |
| Chicago, IL, USA | Chicago, IL, USA | Chicago, IL, USA | Chicago, IL, USA |
| marekniekrasz1@gmail.com | giurcanu@uchicago.edu | tjcarroll@uchicago.edu | gchristofo@yahoo.com |

**Abstract:**
**Objectives**: Quantification of perfusion in ml/100g/min, rather than comparing relative values side-to-side, is critical at the clinical and research level for large longitudinal and multicenter trials. Intravoxel Incoherent Motion (IVIM) is a non-contrast Magnetic Resonance Imaging (MRI) diffusion-based scan that uses a multitude of b-values to measure various speeds of molecular perfusion and diffusion, sidestepping inaccuracy of arterial input functions or bolus kinetics in quantitative imaging. Questions remain as to the signal origin and whether IVIM returns quantitative and accurate perfusion in a setting of pathology. As such, we test a new method of IVIM quantification and compare our values to reference standard neutron capture microspheres across three physiologic states in a controlled animal model.
**Materials and Methods:** We derive an expression for the quantification of capillary blood flow by solving the 3D gaussian probability distribution and defining water transport time as when 50% of the molecules remain in the tissue of interest. Our calculations were verified in a study of six subjects that underwent a two-day controlled experiment in a pre-clinical canine model of normocapnia, CO2 induced hypercapnia, and middle cerebral artery occlusion (ischemic stroke). IVIM perfusion was quantified in ml/100g/min. IVIM water transport time (a surrogate for dynamic susceptibility contrast "Mean Transit Time" (DSC MTT)) was compared to DSC MTT, and IVIM infarct volume was compared to diffusion tensor imaging infarct volumes. Simulations were studied to suppress non-specific cerebrospinal fluid (CSF).
**Results:** Linear regression between MTT and IVIM water transport time asymmetry in infarcted tissue was excellent (slope = .59, intercept = .3, $R^2$ = .93). Strong linear agreement also was found between IVIM and reference standard infarct volume (slope = 1.01, $R^2$ = .79). Simulation of CSF suppression via inversion recovery returned blood signal reduced by 82% from combined T1 and T2 effects. Intra-physiologic state comparison of perfusion shows potential partial volume effects which require further study especially in disease states. Linear regression analysis of IVIM and microsphere perfusion returned correlation (slope = .55, intercept = 52.5, $R^2$ = .64), with a Bland-Altman mean difference of -11.8 ml/100g/min.
**Conclusions:** IVIM infarct volumes and quantitative cerebral perfusion correlated with reference standard values over a range of physiologic conditions when quantified with water transport time. The accuracy and sensitivity of IVIM provides evidence that observed signal changes reflect cytotoxic edema and tissue perfusion. Further, partial volume contamination of CSF may be better removed during post-processing rather than with inversion recovery to avoid artificial loss of blood signal.

**Introduction:**

Intravoxel incoherent motion (IVIM) is a post-processing algorithm for diffusion MRI scan introduced by Le Bihan[1] that uses multiple *b*-values to simultaneously assess tissue water diffusion and tissue perfusion. Prolonged scan times, ease of use, and direct quantification of perfusion afforded by contrast-enhanced scans largely supplanted IVIM as a clinical option. However, since newer, faster scanner hardware and software have substantially reduced scan and image processing times, many groups[2-6] are beginning to revisit IVIM perfusion-diffusion image as a tool for evaluating neurovascular disease. IVIM acquisitions could serve a prominent role in the management of neurovascular disease, particularly when longitudinal imaging is desirable. Furthermore, if cerebral blood flow can be quantified in ml/100g/min (qCBF) without organ-dependent capillary geometry assumptions, IVIM has the potential to allow multi-site cross-sectional comparison in addition to within-patient longitudinal comparison in organs besides the brain[7,8].

There is some question of what IVIM signal changes correspond to, with potential that observed "perfusion" changes are reflecting flow of the perivascular cerebrospinal fluid (CSF) and movement of blood in the capillary bed. This study uses a previously reported canine model of acute ischemia[9-11] to examine if quantified IVIM in ml/100g/min agrees with reference standard quantitative values of perfusion and demonstrates similar sensitivity to physiologic changes (both predicted and directly measured). IVIM cerebral perfusion is compared to neutron capture microsphere deposition, an established reference standard for tissue perfusion, at normocapnia (PaCO2 = 30-35 mmHg), hypercapnia (PaCO2 60-70mmHg), and focal ischemia resulting from middle cerebral artery occlusion (MCAO). IVIM water transport time used to quantify perfusion was compared to the similar dynamic susceptibility (DSC) mean transit time via the central volume principle. Lastly, the accuracy of an IVIM-based infarct volume was compared to the infarct volume determined by standard mean diffusivity (MD) at 2 hours post-occlusion. The model is a clinically viable diagnostic tool for identifying perfusion/diffusion changes and allowed for precise control over physiologic conditions and direct comparison to invasive reference standard values of perfusion.

The IVIM sequence and post-processing presented here presents a clinically feasible scan time of 5-minutes to collect images for calculation of parametric perfusion, diffusion, and water transport time MR images. We hypothesize that IVIM perfusion values can be quantified (ml/100g/min) using water transport time and agree with neutron capture microspheres over a range of physiologic conditions. Using the central volume theorem, IVIM water transport time should correlate with DSC mean transit time. Further, using the same IVIM acquisition we hypothesize that IVIM diffusion positive infarct volumes will be accurate compared to MRI mean diffusivity infarct volumes as a reference standard. The purpose of this study was thus to assess the accuracy and sensitivity of a novel quantification of five-minute simultaneous IVIM perfusion and diffusion across three physiologic states.

**MATERIALS AND METHODS**

**IVIM Quantification via Water Transport Time:** Perfusion and diffusion values are calculated from the multiple b-value IVIM scan using software developed in-house with a two-step fitting algorithm built using previous literature [1,7,8]. In our implementation, the diffusion regime ($b > 222 \ s/mm^2$) is fit to the mono-exponential $(1-f)e^{-Db}$ and the perfusion regime is fit to the remainder $fe^{-D^*b}$, resulting in the following bi-exponential on a voxel-by-voxel basis:

$$\frac{S(b)}{S(0)} = fe^{-D^*b} + (1-f)e^{-Db} \quad (1)$$

In Eq. (1) $D^*$ is pseudo-diffusion coefficient, $f$ is blood volume fraction (CBV) and $D$ is the diffusion coefficient.

With the bi-exponential fit returning an approximation of a pseudo-diffusion coefficient $D^*$, the movement of this perfusing water can be written as a 1D gaussian $\int_{-.5}^{.5} \frac{1}{\sqrt{4\pi D^* t}} e^{-\frac{x^2}{4D^*t}} dx$. "Water transport time" is then defined as the time for concentration of perfusing water in a capillary bed to drop to 50% of its original value. With three directions taken this can be written in spherical coordinates for 3D pseudo-diffusion and solved for $t$ as "water transport time" from Eq. (2)

$$\frac{4\pi}{\left(\sqrt{4\pi D^* t}\right)^3} \int_0^{.5} e^{-\frac{r^2}{4D^*t}} r^2 \, dr = 0.5 \quad (2)$$

An example of this 3D spherical diffusion, the change of concentration over time, and the 'water transport time' at which the concentration is 50% of the original is shown in Figure 1. Quantitative CBF can be calculated by substituting this definition of water transport time into the central volume equation, where water transport time is solved from Eq. (2) as a function of $D^*$ $[mm^2/s]$. Fast-water fraction ($f$) per water transport time ($t$) is proportional to cerebral blood flow in ml/100g/min returning Eq. (3) with assumed values of $\rho = 1.04 g/mL$, water content fraction $f_w = .79$, and converted to 100g/min.

$$qCBF \ [ml/100g/min] = \frac{CBV}{WTT} = \frac{f \times f_w}{\rho} \frac{2D^*}{(.32)^2} \approx fD^* \times 93,000 \quad (3)$$

Using Eq. (3) on IVIM bi-exponential parameters from Eq. (1) returns IVIM maps of qCBF in ml/100g/min for comparison to reference standard values.

**Pre-clinical Model:** All experiments were conducted using a pre-clinical canine model of both controlled hypercapnia and MCAO[9,11]. The two-day experimental protocol was approved by the University of Chicago Institutional Animal Care and Use Committee and reported in compliance with ARRIVE guidelines. The University of Chicago is an AAALAC International accredited institution adhering to the following guidelines, regulations, and policies: a) Guide for the Care and Use of Laboratory Animals (National Research Council), b) USDA Animal Welfare Act and Animal Welfare Regulations, and c) Public Health

Service Policy on Humane Care and Use of Laboratory Animals. The canine model has several advantages compared to rodents as a model for assessment of infarct progression using perfusion and diffusion imaging studies. Canines have a gyrencephalic neocortex with similar ratio of white to grey matter [12-14] as well as comparable pial arteriolar network organization[15], critical in modeling collateral arterial blood supply and predicting infarct evolution[16], to humans. In addition, the canine model provides tissue volumes necessary to evaluate both core and penumbra during ischemia using microsphere deposition. The neurovascular structure of canines also allows a range of endovascular devices and interventional radiology techniques allowing minimal invasion with real-time visualization of occlusion preventing imaging artifacts and traumatic cerebrovascular reaction from open surgical occlusion. Lastly, the middle cerebral artery occlusion proposed is a relatively inexpensive alternative to nonhuman primate models of acute ischemia.

**Experimental Protocol:** In each animal, two experiments were performed on consecutive days to validate our algorithm in a setting of cerebrovascular response to hypercapnia and MCAO and the effect of flow augmentation on collateral arterial networks after MCAO[11,16]. We present an analysis of these experiments in which neutron capture microspheres and IVIM were acquired during quiescent physiologic conditions. On the first day, canines were imaged at baseline (i.e., normocapnia) (target PaCO2 = 30-35 mmHg) and hypercapnia (5-7% carbogen respiration, target PaCO2 60-70mmHg) induced by carbogen gas inhalation (5-7% CO2 95% O2). End Tidal CO2 was monitored to assess minute to minute variations in exhaled CO2 levels, and true arterial paCO2 was measured through direct arterial blood sampled from the descending aorta at the time of perfusion measurements.

On the second day, canines underwent permanent endovascular middle cerebral artery occlusion under fluoroscopic guidance. Occlusion was then verified via selective ipsilateral and contralateral internal carotid and vertebral arteriography[17]. Subjects were transported to the MRI suite for imaging studies within 60 minutes of occlusion. We evaluated experimental flow augmentation during MCAO via simultaneous administration of a vasopressor and vasodilator intended to augment CBF[10]; therefore, highly dynamic, and abnormally high perfusion throughout the brain post-occlusion is expected by experimental design. Throughout experiments physiologic parameters were maintained within normal range (excluding PaCO2 for experiments involving intentional hypercapnia and blood pressure for experiments involving flow augmentation).

IVIM and microspheres included in the study were taken less than 30 minutes apart on day two post-occlusion to mitigate perfusion fluctuation. Furthermore, physiologic parameter variation (heart rate, blood pressure, ETCO2) between microsphere and IVIM imaging was monitored. Large variance in physiologic parameters within these timing windows is noted.

**Microsphere Acquisition:** Stable-isotope microsphere CBF in ml/100g/min was acquired using neutron activation[17]. Each injection consisted of 4ml of stable isotope labeled 15μm microspheres (STERIspheres,

BioPal Inc, Medford, MA) into the left ventricle over 10s and reference blood (20ml) was withdrawn at 10ml/min for analysis. Brains were excised and sectioned into regions of interest (ROIs), Figure 2, and analyzed through neutron activation at an independent laboratory (BioPal Inc. Medford, MA, USA).

**MRI Acquisition:** All MRI scans were performed on a 3T MRI scanner (Ingenia, Philips) with canines in a headfirst, prone position using a 15 channel receive-only coil. Diffusion weighted images (DWI) for IVIM were collected with 10 b-values from 0 to 1000 s/mm$^2$ (0,111, 222, 333, 444, 556, 667, 778, 889, 1000) and 3 orthogonal directions to ensure a clinically feasible scan time (5:38 min). IVIM scans were prescribed to cover the entire head (2D single shot EPI, TR/TE= 3056/91ms, 50 slices, 2mm, FOV = 160 mm, total scan time = 332s, SENSE Factor=2). Dynamic Susceptibility Contrast (DSC) images were acquired using a T1-Bookend method[17] within 30 minutes of IVIM (FOV/Matrix = 220 mm/224, single shot, EPI, Fat Saturated, No. of slices = 5, Slice Thickness = 6 mm, TR/TE = 315/40, Flip angle = 75°, 200 time points) and delay and dispersion corrected[18] to return Mean Transit Time (MTT). Diffusion tensor imaging (DTI) for infarct volume progression was prescribed to cover the entire head (TR/TE = 2993/83ms, FA = 90°, BW = 1790Hz/pixel, FOV =160 mm, voxel size = $1.75 \times 1.75 \times 2\ mm^3$, b-values = 0, 800 s/$mm^2$, slices = 50, 32 directions).

**Perfusion Comparison:** Hemispheric regions-of-interest (ROIs), shown in Figure 2, were manually drawn on three central slices on the diffusion image by a trained operator who was blinded to all perfusion values for comparison to microsphere perfusion. IVIM slices were averaged to match microsphere thickness (6mm). To remove false values from misfitting, $D^* \geq .10$ were removed. Manual avoidance of subarachnoid space and ventricles, along with leave-one-out cross validation T2 map thresholds, were used to remove CSF-dominated voxels from analysis. Kwong et al.[19] suggests using an inversion recovery saturation pulse to remove proposed artifactual high perfusion values in the brain from the glymphatic system. In this study a simulation was written to explore the effects of this saturation pulse on blood signal.

**Transit Time Comparison:** Anatomic middle cortical ROIs post-MCAO were drawn by an operator blinded to physiologic status on pre-processed images of IVIM and DSC ipsilateral and contralateral to the occlusion.

**Infarct Volume Comparison:** Mean Diffusivity (MD) analogous to more widely used Apparent Diffusion Coefficient (ADC), was used to determine true infarct volume with the automatic threshold of 5.7e-4[9]. A previously determined automatic IVIM $D$ infarct threshold of 5.15e-4 was applied calculate IVIM infarct volume.

**Statistical Analysis:** Linear regression was analyzed, pooled across all physiologic states and the coefficient of determination ($R^2$), p-values (*p*), slopes (*m*), offsets (*b*), and corresponding Bland-Altman mean differences and 95% CIs are reported. Tukey's boxplots per physiologic state with mean, first and third quartiles, and whiskers were produced with hemispheric paired comparison. Statistical agreement within physiologic states was studied with Wilcoxon signed-rank. Lin's concordance correlation coefficient (CCC) was used to determine agreement between multiple quantitative measures incorporating both accuracy and precision between two readings along the 45 degree line through the origin[20]. In addition, CCCs variability across subjects was shown with the reported subject-wise CCC. All statistical significance was determined at the 1% ($p < .01$) level for hypothesis testing.

**RESULTS:** A total of n=6 (mean age = .72y, mean weight = 25.1kg, 5 female, 1 male) experiments were analyzed with further detail in Table 1. Representative images are shown in Figure 3A) for quantitative perfusion images during normocapnia, hypercapnia and post-MCAO, in Figure 3B) for transit time post-MCAO, and in Figure 3C) for diffusion-positive maps post-MCAO.

Figure 4A-B) shows linear regression, Lin's CCC, and corresponding Bland-Altman of IVIM and microsphere perfusion. Tukey's Box Plots in Figure 4C) demonstrate average and range of perfusion at normocapnia, hypercapnia, and post-MCAO split into ipsilateral and contralateral hemispheres. There were no significant differences in hypercapnia or post-occlusion hemispheres, but a statistically significant difference was seen at normocapnia (p=0.005). Linear regression of middle cortical asymmetry of IVIM water transport time post-occlusion returned correlation to DSC mean transit time (Figure 5A) and Tukey's Box plots show expected increase in transit time on the ipsilateral hemispheres, though with a different mean and range of values (Figure 5B). Automatic infarct volume from IVIM returned strong linear correlation to MD ($slope = 1.01, intercept = .58, R^2 = .71, CCC = .79$). Subject-wise Lin's CCC for cases with all slices analyzed ranged from .26 to .86, with a mean of .56. Simulation of inversion recovery spin echo to suppress CSF signal in IVIM showed the optimal TI for suppression would also suppress blood signal to 18% of its original (Figure 6).

**DISCUSSION**

This study found that a 5-minute IVIM 10 b-value sequence can produce quantitative cerebral perfusion via water transport time in a clinically acceptable scan time across a range of relevant physiologic conditions. The sensitivity of CBF to normocapnia, hypercapnia, and acute ischemia post-MCAO suggest IVIM images can have impact in the management of cerebrovascular disease. Furthermore, IVIM images can predict qCBF in response to controlled changes in CO2 (cerebrovascular reactivity), without repeated administration of contrast agent, an advantage in the evaluation of chronic disease[7,21]. Further, the study suggests a possible quantification factor that relies solely on pseudo-diffusion water transport, and not on organ-dependent capillary geometry. All comparisons between IVIM and microspheres across three physiologic states returned linear regression correlation and were further supported by agreement shown with Lin's CCC analysis.

Prior studies compare IVIM cerebral perfusion to MR CBF measurements with varying levels of success[7,22,23]. Several studies compare DSC CBV and IVIM $f$ [7,22], while one study ($R^2 = 0.42$) compares relative IVIM CBF $fD*$ and relative DSC CBF[23]. In comparison, this study found a strong correlation ($R^2 = 0.64$) of quantitative IVIM CBF $fD*$ in ml/100g/min to neutron capture microsphere qCBF in ml/100g/min. As linear regression suggests sensitivity across three physiologic states, paired agreement within physiologic states supports quantification of hemispheric perfusion without capillary geometry assumptions in a multi-day controlled pre-clinical model which to the best of our knowledge has not been published previously. However, it also highlights an example of the remaining difficulty of CSF removal: partial volume contamination of CSF may falsely add to perfusion signal at normocapnia shown in Figure 4C) and with $p < .01$ at normocapnia.

Water transport time used to quantitate perfusion showed linear correlation with delay and dispersion corrected DSC mean transit time asymmetry in the affected middle cortical territory post-MCA occlusion. Comparison of IVIM WTT to DSC requires DSC undergo delay and dispersion correction, i.e. a Local AIF, to account for IVIM AIF-independence. Without this correction of DSC, correlation of qCBF and transit time would be weaker. Difference in transit time range observed is notable (Figure 4B), likely due to the difference in properties being measured and endogenous vs. exogenous measurement.

Previous studies have also reported positive correlation between ADC and IVIM $D$ in ischemic and normal regions ($R^2 = 1, R^2 = 1$) [24] and ($R^2 = .98, R^2 = .81$) [25]. Although this study's agreement was lower than prior studies (slope = 1.01, $R^2 = .79$), it should be noted that the IVIM and DTI scans that were compared were taken from separate MR scans during a period of dynamic infarct expansion while the previous studies used the same images to get ADC and IVIM D. Use of two separate scans and sequences for IVIM D and MD infarct volume was chosen to best compare IVIM D and a separate standard clinical infarct volume. This combined with the IVIM qCBF from the same IVIM scan, not included in previous studies, demonstrates that the simultaneous perfusion and diffusion components from a single IVIM bi-exponential fit agreed with corresponding perfusion and diffusion reference standard values.

IVIM imaging exploits the fact that diffusion-weighting is sensitive to water motion and consequently able identify where in the brain water is sequestered inside cells by cytotoxic edema.[26] At very low b-values ($b \leq 200 \ s/mm^2$), diffusion-weighting is predominantly sensitive to faster water motion such as capillary level blood flow[8,21,27-29]. Questions remain regarding the origin of fast water motion signal, with Kwong et al.[19] showing that use of an inversion recovery pulse significantly reduced cortical gray matter pseudo-diffusion fraction suggesting that much of the bi-exponential behavior resulted from CSF contamination rather than perfusion[19]. Figure 6 shows inversion recovery spin echo suppression will suppress both CSF and blood meaning signal loss with CSF suppression may come from inversion recovery reducing signal from both CSF and blood from both T1 and T2 effects. Further, following the Monro-Kellie hypothesis, at hypercapnia the dilation of cerebral capillaries prompts an increase in cerebral blood volume and CBF, and a decrease in CSF to maintain intracranial pressure[30]. If fast-flow IVIM signal were due solely to CSF contamination, $fD^*$ should decrease after CO2 inhalation, which was not observed in this study (Fig. 4C). Use of an absolute T2 threshold to remove CSF-dominated voxels in the ventricles and subarachnoid space reduced the effect of human bias in region selection. However, endogenous contrast prevents proof of signal origin and partial volume effects require further study especially in disease states. This study provides evidence that observed IVIM signal changes reflect cytotoxic edema and tissue perfusion.

Arterial spin labelling (ASL) is well-established, widely available, and fully quantitative means of imaging cerebral perfusion noninvasively [31]. However, with ASL, delayed arterial arrival times due to proximal vessel occlusion in stroke studies remains a challenge [32,33]. As such, IVIM being independent of arterial input functions or bolus kinetics is a singificant benefit of the sequence in comparison to both ASL and DSC-MRI when measuring neurovascular disease. Futher, IVIM is able to provide simultaneous identification of cytotoxic edema unlike both ASL and DSC-MRI.

This study is not without limitations. The highly dynamic physiology created challenges for acquisition of normocapnia, hypercapnia, and post-MCAO with MRI and invasive neutron capture microspheres within tight timing windows. Two cases showing large disagreement between IVIM and microsphere perfusion also show rapid change in BP and Co2 that agrees with the discrepancy shown (Fig 4B), though the causality cannot be proven. While sensitivity to physiologic change is observed, IVIM sensitivity appears weaker than the perfusion effects observed with microspheres. Notably normocapnia returns higher IVIM perfusion, possibly due to partial volume contamination from CSF. Exponential fitting error especially regarding pseudo-diffusion coefficient is a continuing problem. While limits were placed to maintain pseudo-diffusion coefficients in the expected range, pseudo-diffusion error will have large effects due to the inverse relation to water transport time. Further, some data was lost beyond our control due to vendor returned zeros and dynamic physiology, highlighting the complexity of microsphere perfusion. Finally, translatability of this perfusion to humans is not known.

**CONCLUSIONS**

In conclusion diffusion positive volumes and quantitative cerebral perfusion agreed with reference standard values over a range of physiologic conditions. The sensitivity of IVIM provides evidence that observed signal changes reflect cytotoxic edema and tissue perfusion. IVIM images were acquired from five-minutes of non-contrast diffusion weighted images and quantified with water transport time independent of capillary geometric assumptions and returned agreement to neutron capture microspheres across a wide range of physiologic states such as normocapnia, hypercapnia, and infarction. This supports the further development and refinement of IVIM for measuring noninvasive quantitative perfusion and diffusion.

**Figures and Tables.**

|  | Normocapnia | Hypercapnia | Occlusion |
|---|---|---|---|
| **IVIM & Microspheres** | 5 (Case 1 IVIM incomplete) | 5 (Case 4 IVIM incomplete) | 5 (Case 5 subarachnoid hemorrhage) |

Table 1. Of the 6 cases studied, shown is the number of cases available for each physiologic states, with all measurements within 30 minutes of each other. Two cases post-MCAO had slices removed from analysis due to zeros returned by the microsphere vendors.

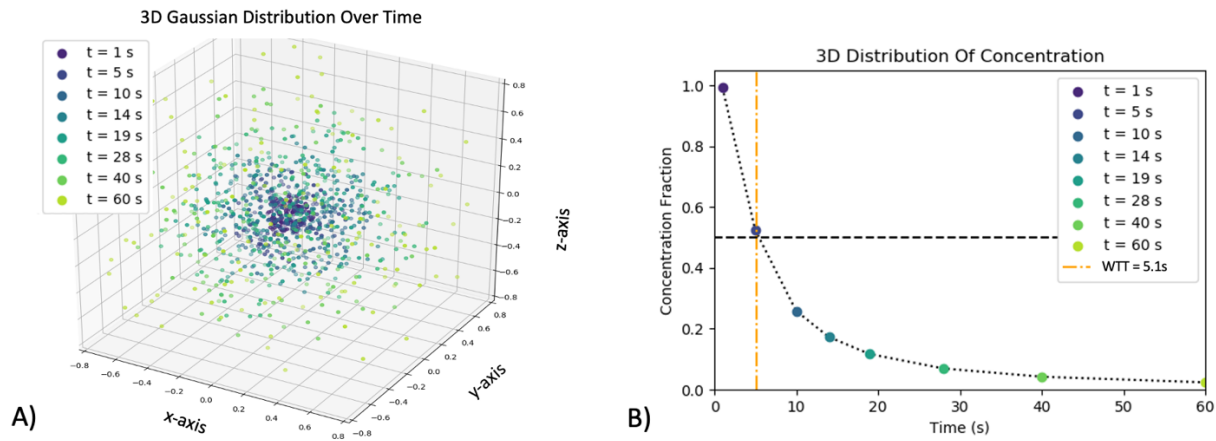

Figure 1. A) A simulation of 3D diffusion for a voxel with a pseudo-diffusion coefficient of $D^* = .009$ with the distribution plotted as it spreads over time. B) is the corresponding concentration as it drops over time, with the orange line showing the water transport time at which the concentration is 50% of the original. This is an example of the transit time this study uses to quantify IVIM perfusion using the central volume principle.

**Representative ROIs**

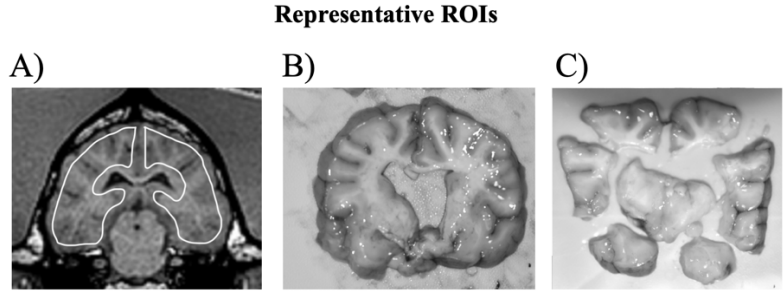

Figure 2. Representative images for hemispheric ROIs on one of three slices drawn on IVIM and on DSC on an Anatomic T1 weighted MRI (A). Edges are avoided to avoid subarachnoid CSF and DSC susceptibility artifacts. The brain slice (B) is cut into eight regions (C) for microsphere analysis and averaged for hemispheric qCBF.

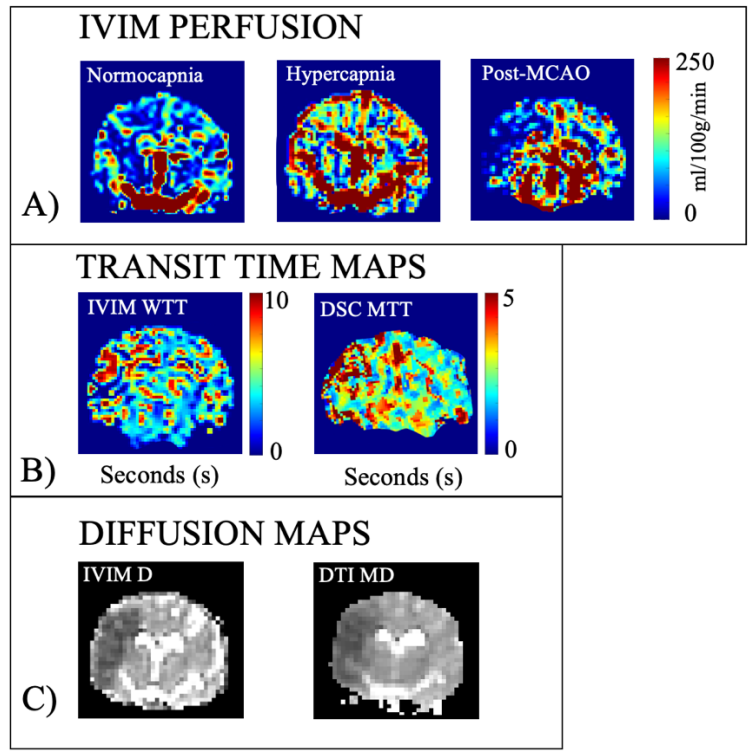

Figure 3. A) Representative quantitative IVIM perfusion for normocapnia, hypercapnia, and post-MCAO respectively. The qCBF images are shown with a dynamic range of 0 to 250 ml/100g/min. B) Transit time maps post-MCAO for (right) delay and dispersion corrected DSC MTT and (right) IVIM water transport time. C) Diffusion images with a Mean Diffusivity map (right) compared to IVIM diffusion coefficient image (left) post-MCAO used for calculation of infarct volume. A case without flow augmentation treatment is provided to highlight diffusion abnormality post-MCAO.

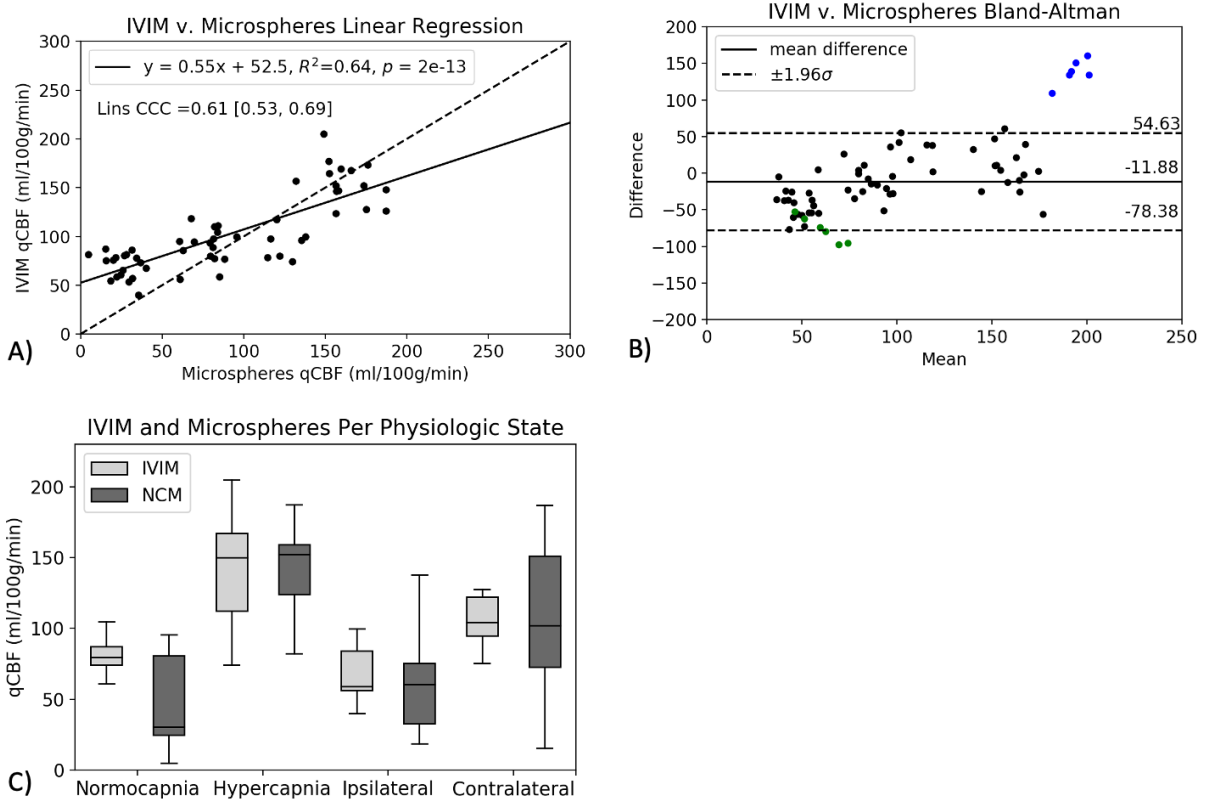

Figure 4. A) Hemispheric correlation of microsphere qCBF vs. quantitative IVIM qCBF across all physiologic states. Solid lines in linear regression plots represent the line of regression and a dashed line of unity is shown for reference. B) Bland-Altman plot with mean difference as a solid line and dashed lines representing 95% CI. Blue and green color represent two cases with large change in BP and CO2 between IVIM and microsphere injection despite being within 30 minutes of acquisition. Due to the instability in physiology these cases are not included in statistical analysis but are shown as outliers in Bland-Altman for the sake of completeness. C) Tukey's box-whisker plots of hemispheric values for all physiologic states. Microspheres are called 'NCM' for brevity. Normocapnia and hypercapnia include both left and right hemisphere while occlusion and contralateral are split into corresponding hemispheres. Note values are considerably higher in the MCAO model due to the use of aggressive flow augmentation[16].

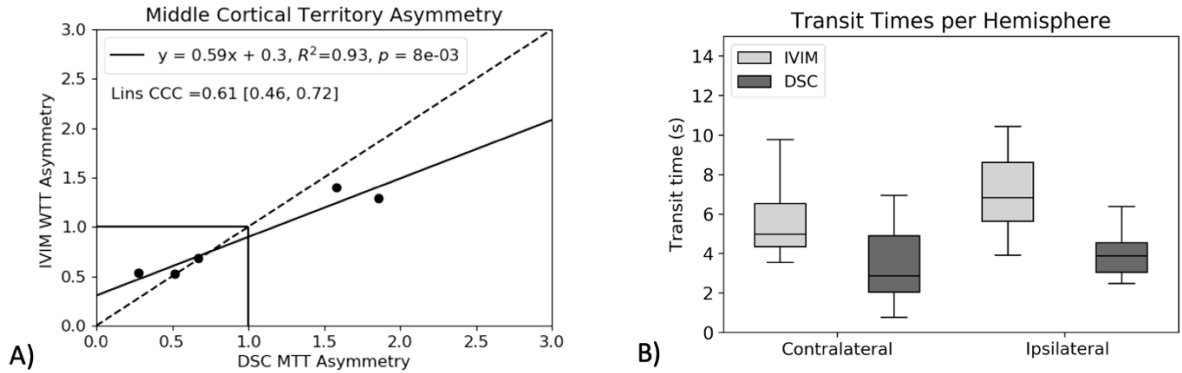

Figure 5. A) Correlation of middle cortical territory asymmetry (Left/Right) of the five subjects post-MCAO. B) Tukey's box blots of the middle cortical territories post-MCAO showing mean and range of transit times from IVIM and DSC, split into contralateral and ipsilateral hemispheres.

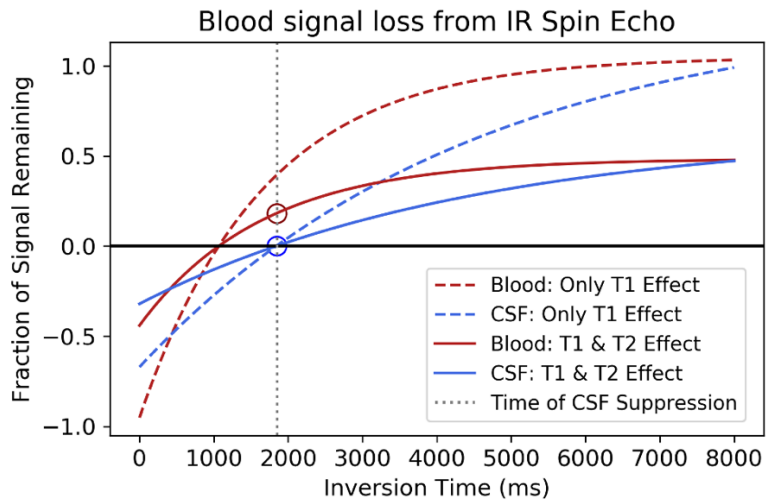

Figure 6. Inversion Recovery Spin Echo suppression simulation (TR=5000ms, TE=37ms) of remaining transverse signal from CSF (blue) and blood (red) after an IR pulse with various inversion times taking both T1 and T2 literature value effects into account. A TI = 1850ms (vertical dotted line) was found to be the best time for CSF suppression (blue circle), at the cost of blood signal suppressed to 18% of original signal (red circle). While only T1 effects (dashed) suggest moderate regrowth of blood signal, when considering both T1 and T2 (solid) blood signal is not recovered (red circle).

**FUNDING**
This study was funded by National Institute of Neurological Disorders and Stroke (R01- NS093908), National Science Foundation (DGE-1746045), American Heart Association (GRNT-20380798).

**APPENDIX**

Step-by-step derivation of water transport time. Beginning with 1D for a simpler thought experiment, a diffusing particle is constrained to a region on a line with a molecule starting in the middle of the line and diffusing. If particles begin as a delta function in the center of this 1D line, the way tracer kinetics is modeled in DSC, following the law of diffusion the distribution of particles will change over time according to a normal distribution as a function of $\sigma = \sqrt{2Dt}$. In 1D this can be solved

$$\int_{-.5}^{.5} \frac{1}{\sqrt{2\pi}\sigma} e^{-\frac{x^2}{2\sigma^2}} dx = .50$$

$$.6745\sigma = .6745\sqrt{2Dt} = .50$$

$$WTT = \frac{(.741)^2}{2D}$$

In other words, for an ensemble of molecules diffusing along a 1D line with diffusion coefficient *D* over time *t*, half of them will have moved further than 1mm and half will not. This is the 1D water transport time. Moving to 3D which is shown in Fig. 1 we get the following:

$$\int_{-.5}^{.5}\int_{-.5}^{.5}\int_{-.5}^{.5} \frac{1}{\sqrt{(2\pi)^3(2Dt)^3}} e^{-\frac{x^2}{4D_x t}} e^{-\frac{y^2}{4D_y t}} e^{-\frac{z^2}{4D_z t}} = .50$$

In spherical coordinates for isotropic 3D diffusion this can be written as

$$\int_0^{2\pi} \int_0^{\pi} \int_0^{.5} \frac{1}{\sqrt{(2\pi)^3(2Dt)^3}} e^{-\frac{r^2}{4Dt}} r^2 \sin\theta \, dr \, d\theta \, d\phi = .50$$

$$\frac{4\pi}{\left(\sqrt{4\pi Dt}\right)^3} \int_0^{.5} e^{-\frac{r^2}{4Dt}} r^2 \, dr = .50 \quad (1)$$

The concentration over time will decrease as water diffuses out of the sphere. Solving for the time at which this concentration is 50%, i.e. the WTT, is show below. For simplicity $s$ is substituted for $4Dt$, and the integral is solved by integration by parts.

$$\int e^{-\frac{r^2}{s}} r^2 \, dr$$

$$\int u \, dv = uv - \int v \, du$$

$$u = r, \quad du = dr$$

$$v = -\frac{s}{2} e^{-\frac{r^2}{s}}, \quad dv = -\frac{s}{2} \left(-\frac{2r}{s} e^{-\frac{r^2}{s}}\right) = re^{-\frac{r^2}{s}}$$

$$= -r \left(\frac{s}{2} e^{-\frac{r^2}{s}}\right) + \int \frac{s}{2} e^{-\frac{r^2}{s}} \, dr$$

Putting in the limits of integration from 0 to .5 returns the following:

$$\int_0^{.5} e^{-\frac{r^2}{s}} r^2 \, dr$$

$$= -r \left(\frac{s}{2} e^{-\frac{r^2}{s}}\right)\Big|_0^{.5} + \int_0^{.5} \frac{s}{2} e^{-\frac{r^2}{s}} \, dr$$

$$= -.5 \left(\frac{s}{2} e^{-\frac{.5^2}{s}}\right) + \frac{s}{2} \frac{\sqrt{\pi s}}{2} \operatorname{erf}\left(\frac{r}{\sqrt{s}}\right)\Big|_0^{.5}$$

$$= -.25s \left(e^{-\frac{.25}{s}}\right) + \frac{\sqrt{\pi}}{4} s^{\frac{3}{2}} \operatorname{erf}\left(\frac{.5}{\sqrt{s}}\right)$$

Plugging this back into Eq. 1 returns:

$$\frac{4\pi}{(\pi s)^{\frac{3}{2}}} \left[-.25s \left(e^{-\frac{.25}{s}}\right) + \frac{\sqrt{\pi}}{4} s^{\frac{3}{2}} \operatorname{erf}\left(\frac{.5}{\sqrt{s}}\right)\right] = 0.5$$

$$-\frac{1}{\sqrt{\pi s}} e^{-\frac{.25}{s}} + \operatorname{erf}\left(\frac{.5}{\sqrt{s}}\right) = 0.5$$

Due to the nature of erf($x$) this is not analytically solvable. Therefore, use of a numerical approximation of erf($x$) is required. The approximation of erf($x$) from Eq. 7.1.27 of Handbook of Mathematical Functions with Formulas, Graphs, and Mathematical Tables[34] returns an error $\leq 5e^{-4}$ and is as follows:

$$\text{erf}(x) = \frac{2}{\sqrt{\pi}} \int_0^x e^{-t^2} dt = 1 - (1 + .278393x + .230389x^2 + .000972x^3 + .078108x^4)^{-4}$$

Plugging this in returns the following equation which can be solved for $s$.

$$-\frac{1}{\sqrt{\pi s}} e^{-\frac{.25}{s}} + \left[1 - \left(1 + .278393\left(\frac{.5}{\sqrt{s}}\right) + .230389\left(\frac{.5}{\sqrt{s}}\right)^2 \right. \right.$$
$$\left. \left. + .000972\left(\frac{.5}{\sqrt{s}}\right)^3 + .078108\left(\frac{.5}{\sqrt{s}}\right)^4\right)^{-4}\right] = 0.5$$

$$s \approx .21$$

This can be checked by plugging $4Dt = s \approx .21$ into Eq. 1 as follows:

$$\frac{4\pi}{(\sqrt{.21\pi})^3} \int_0^{.5} e^{-\frac{r^2}{.21}} r^2 \, dr = .50281$$

Returning to Gaussian diffusion parameters if we substitute $\sigma = \sqrt{2Dt} = \frac{\sqrt{s}}{\sqrt{2}}$ we can solve for $\sigma = \frac{\sqrt{.21}}{\sqrt{2}}$, and get $\sigma \approx .32$.

Solving this for t in terms of D returns:

$$WTT = \frac{(.32)^2}{2D}$$

This is the same as if we solved for $t$ in terms of $D$ straight from the approximation of $s$, but in terms of $\sigma$ of the the 3D Gaussian for easier comparison to an intuitive 1D. With this water transport time derived and explained intuitively, we can plug this in to the central volume equation to return the calibration coefficient for the fast-moving component of IVIM volume $f$ and pseudo-diffusion coefficient $D^*$.

$$qCBF = \frac{CBV}{WTT} = \frac{f \times f_w}{\rho} \frac{2D^*}{(.32)^2}$$

It should be emphasized that this is an approximation. The water content fraction of a voxel $f_w$, the density $\rho$, the numerical estimation used for solving the erf($x$), and the assumption of it being a sphere all contribute to the quantification factor above. This is the tradeoff in assumptions compared to capillary bed network assumptions. Using a WTT based perfusion model:

$$qCBF_{WTT} = fD^* \times \frac{2f_w}{\rho(.32)^2}$$

$$= fD^* \times \frac{2(.79)}{1.0\frac{g}{mL}(.32)^2} \times 100g \times 60\frac{s}{min}$$

$$\approx fD^* \times 93000$$